\documentclass[preprint,aps,prb]{revtex4}
\usepackage{graphicx}
\usepackage{dcolumn}
\usepackage{bm}
\begin{document}
\title{Magnetic Splitting of the Zero Bias Peak in a Quantum Point Contact with a Tunable Aspect Ratio}
\author{Tai-Min Liu} \affiliation{Department of Physics, University of Cincinnati, Cincinnati, Ohio 45221, USA}
\author{Bryan Hemingway} \affiliation{Department of Physics, University of Cincinnati, Cincinnati, Ohio 45221, USA}
\author{Steven Herbert}
\affiliation{Physics Department, Xavier University, Cincinnati, Ohio 45207, USA}
\author{Michael Melloch}
\affiliation{School of Electrical and Computer Engineering, Purdue University, West Lafayette, Indiana 47907, USA}
\author{Andrei Kogan}
\email{andrei.kogan@uc.edu}
\affiliation{Department of Physics, University of Cincinnati, Cincinnati, Ohio 45221, USA}

\begin{abstract}
We report a zero-bias peak (ZBP) in the differential conductance of a Quantum Point Contact (QPC), which splits in an external magnetic field. The peak is observed over a range of device conductance values starting significantly below $2e^2/h$. The observed splitting closely matches the Zeeman energy and shows very little dependence on gate voltage, suggesting that the mechanism responsible for the formation of the peak involves electron spin.  Precision Zeeman energy data for the experiment are obtained from a separately patterned single-electron transistor located a short distance away from the QPC. The QPC device has four gates arranged in a way that permits tuning of the longitudinal potential, and is fabricated in a GaAs/AlGaAs heterostructure containing two-dimensional electron gas. We show that the agreement between the peak splitting and the Zeeman energy is robust with respect to moderate distortions of the QPC potential. We also show that the mechanism that leads to the formation of the ZBP is different from the conventional Kondo effect found in quantum dots.
\end{abstract}

\maketitle

Current-voltage characteristics of ballistic quantum point contacts (QPCs) (Ref. \onlinecite{wharam:88}) -- narrow channels contacted by macroscopic conductors -- have been proven difficult to understand despite the geometric simplicity of the QPC devices. Ballistic flow of electrons in QPCs produces plateaus in the linear conductance $G$   separated by $2e^2/h$, which are found in many experiments  and  are well understood \cite{vanWees:88}. Yet, the nonlinear $I-V$ characteristics of  many QPCs show a zero-bias peak (ZBP), which is challenging to explain by single-particle effects alone. In this report, we focus on the properties of the ZBP at relatively low conductance values, $<0.5 e^2/h$. This regime differs from the near-opening regime,  in which the so-called  ``0.7 anomaly" in the linear conductance is often observed \cite{Thomas:96,Kristensen:00,cronenwett:02,koop:07} and where the effect of spin correlations on nonlinear transport has been of significant interest \cite{cronenwett:02,Rejec:06,luscher:07,sfigakis:08}. Recent experiments \cite{luscher:07} and theory \cite{Rejec:06} suggested that, near the opening,  localization of unpaired spins in QPCs may occur and produce  a ZBP due to an analog of the Kondo effect \cite{cronenwett:02,Meir:02}. At the same time, an interpretation of the ZBP that does not involve electron spin was recently proposed \cite{chen:09}. In this paper we show that a ZBP related to electron spin can occur at conductance values significantly lower than $0.7\times  2e^2/h$, and this does not involve the conventional spin 1/2 Kondo effect.

Our ZBP measurements are obtained with a semiconductor QPC sample  that has 4 independent gates, which we use to manipulate the device potential profile along the direction of the  flow of the QPC current. When no significant distortion of the potential is present, we find a clear ZBP at conductance values substantially below the first plateau. The ZBP splits with the application of an in-plane magnetic field $B$, applied perpendicular to the current flow direction.  Further, we show that the result is robust against moderate distortions of the longitudinal potential. Distorting the potential by a large amount, however, produces a real bound state, likely localized between the device gates, as evidenced by the characteristic Coulomb blockade (CB) diamond and a zero-bias peak that we attribute to the conventional  Kondo physics found in quantum dots. Importantly, when the QPC potential is ``smooth" and  the CB is not observable, the ZBP is still present and shows clear splitting with the magnetic field applied. The splitting closely matches precision Zeeman energy, defined as $\Delta_Z= g^*\mu_B B$, where $g^*$ is the effective electron g-factor, $\mu_B$ is the Bohr magneton and $B$ is field, which we obtain independently from a single-electron transistor (SET) on the same chip. This  shows that the ZBP in this regime is still related to the electronic spin, and  rules out  the conventional Kondo physics due to an accidental trapping of an unpaired electron in the device, which would produce CB features in addition to the ZBP.

\begin{figure}
\includegraphics[width=5in, keepaspectratio=true]{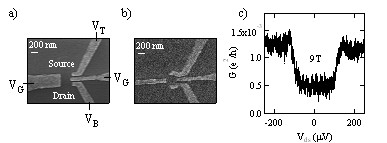}
\caption{\label{fig:dev} (a) Micrograph of a four-gate QPC nominally identical to that used in measurements with the gate voltage labeling convention shown. (b) Micrograph of an SET placed $\sim 150$ $\mu$m away from the QPC device on the same chip for Zeeman energy measurement. (c) The plot of the nonlinear conductance of the SET device in the spin-flip cotunneling regime showing characteristic steps at $V_{ds}=\pm \Delta_{Z}/e$. The step-to-step width is two times of the Zeeman splitting $\Delta_{Z}/e$.}
\end{figure}

The differential conductance $G=d I / d V_{ds}$ of our QPC is measured via standard lock-in techniques with the excitation voltage of approximately 3.9 $\mu$V RMS at 17 Hz. The four gates of our QPC are  arranged on top of a GaAs/AlGaAs heterostructure (electron sheet density  $n_{2D}=4.8\times 10^{11}$ cm$^{-2}$ and mobility $\mu=5\times 10^{5}$ cm$^2$/V\,sec at 4.2K) as shown in Fig. \ref{fig:dev}(a). The same voltage $V_G$ is applied to the two opposing gates, and the voltages $V_T$ and $V_B$ can be tuned to adjust the longitudinal potential profile. A nearby SET device [Fig. \ref{fig:dev}(b)], patterned approximately 150 $\mu$m away from the QPC, is used to measure the Zeeman energy and the electron temperature via spin-flip cotunneling spectroscopy \cite{Kogan:04,Zumbuhl:04,Loss:06}. In this regime, the conductance of the SET shows steps at $V_{ds}=\pm \Delta_{Z}/e$ [Fig. \ref{fig:dev}(c)] and the steps slopes can be used to obtain the electron temperature, about 55 mK in our devices, as described in Refs. \onlinecite{Kogan:04} and \onlinecite{liu:09}. Fitting the Zeeman energy linearly with the magnetic field gives the exact heterostructure g-factor $|g|=0.2073\pm 0.0013$ \cite{liu:09}, which is much smaller than the bulk GaAs value $|g|=$0.44.

\begin{figure}
\includegraphics[width=5in, keepaspectratio=true]{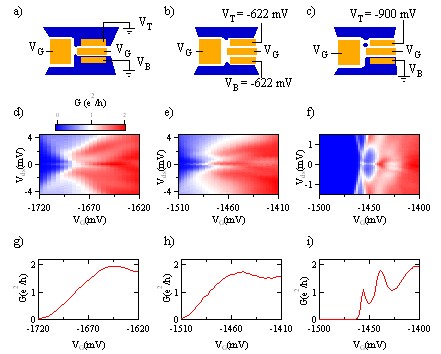}
\caption{\label{fig:data} (Color) (a-c) Three representative device settings  discussed in the text. Areas accessible to electrons are schematically shown by blue color.  (a) Short constriction, with $V_T$ and $V_B$ both set to zero. (b) Long constriction, with $V_T$ and $V_B$ both at a negative bias. (c) A strongly distorted potential, resulting in a formation of a quantum dot between the gates.   (d-f) Nonlinear conductance maps for the three regimes shown above. A Coulomb diamond, clearly seen in (f), is not present in either (d) nor (e). The ZBP is seen in each dataset. (g-i) Linear conductance data corresponding to the plots shown in (d) and (e). (i) An onset of Coulomb oscillations signaling a formation of a quantum dot in the constriction. No such oscillations are present in (g) or (h).}
\end{figure}

\begin{figure}
\includegraphics[width=5in, keepaspectratio=true]{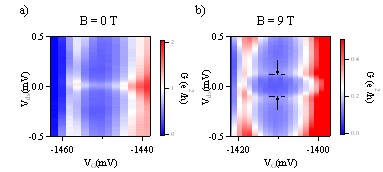}
\caption{\label{fig:splitting_kondo} (Color) Non-linear conductance data for the  constriction in a Coulomb blockade regime with the gate voltages set as shown in Fig. 2(c). (a) The portion of the data shown on Fig. 2(i) corresponding to the CB diamond region at zero magnetic field. (b) The same gate scan as in (a)  with a 9 T in-plane magnetic field present. The horizontal lines marked by the arrows show the Zeeman bias voltage threshold.}
\end{figure}

Figure \ref{fig:data} shows three representative gate voltage settings (a)--(c) and the corresponding nonlinear conductance maps (d)--(f) and the zero bias conductance curves (g)--(i). In each presented measurement, the voltage $V_G$, applied to the opposing center contacts, is scanned. The voltages $V_T$ and $V_B$ applied to the top and the bottom gates, respectively, control the longitudinal potential profile: by setting both $V_T$ and  $V_B$ to zero, as shown in Fig. \ref{fig:data}(a), we produce a ``short" constriction, formed by the center gate alone. Applying a moderate negative voltage to $V_T$ and $V_B$ [Fig. \ref{fig:data}(b)] increases the  constriction length. We note that, in both regimes, a zero-bias peak in the nonlinear conductance is observed over a range of values of $V_G$, and the linear conductance rises to a value close to $2 e^2/h$ monotonically. As expected, the pinch-off voltage  for the $V_T=V_B=0$ data  is more negative than for the $V_T=V_B=-662$ mV data. An example of a strong distortion of the potential is shown in Fig.  \ref{fig:data}(c). A dramatic change in the conductance properties is observed in this regime: a characteristic ``diamond" appears in the nonlinear conductance plot [Fig. \ref{fig:data}(f)], and the linear conductance displays sharp peaks before the $2 e^2/h$ plateau is reached, both typical features of a quantum dot in a CB regime \cite{Kouwenhoven:97}. We attribute this behavior to a quantum dot  forming between the electric field fringes created by the middle and the top gates, as shown in Fig. \ref{fig:dev}(c).  Importantly, a zero-bias peak is present across the diamond [Fig. \ref{fig:data}(f)], which splits when an in-plane magnetic field is applied [Fig. \ref{fig:splitting_kondo}(b)]. We interpret this feature as the conventional Kondo effect often observed in quantum dots. With the strength to control the device regime, we thus avoid this Kondo feature to investigate the ZBP in QPCs.

\begin{figure}
\includegraphics[width=5in, keepaspectratio=true]{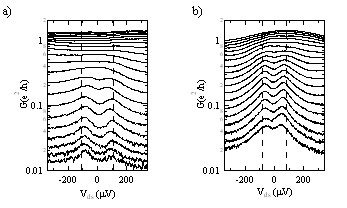}
\caption{\label{fig:data_vs_gate} (a) and (b) Observed ZBPs for the short constriction: $V_G$ from -1507 (bottom) to -1470 mV  and long constriction: $V_G$ from -1634 (bottom) to -1593 mV (top) at zero magnetic field. Data shown in the figure are obtained by scanning $V_G$ only. (c) and (d) Magnetic field dependence of the peak shape for the two configurations, showing that the splitting increases with the field in both cases. The traces are taken at $V_G$=-1630 mV (short) and $V_G$=-1490 mV (long).}
\end{figure}

\begin{figure}
\includegraphics[width=5in, keepaspectratio=true]{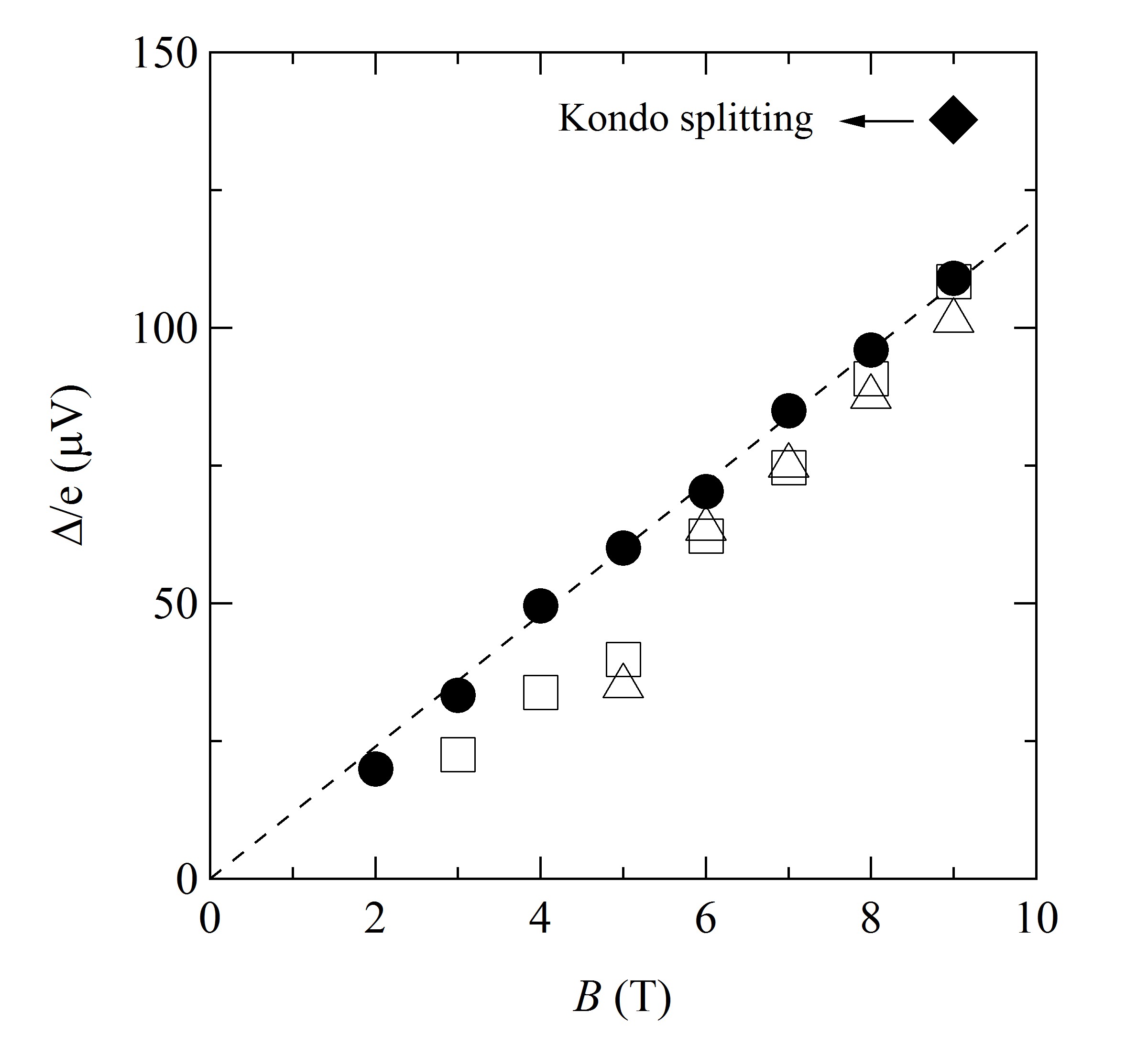}
\caption{\label{fig:split} Comparison between the Zeeman energy, ZBP splitting in different regimes, and Kondo splitting. (Filled circles) Zeeman splitting obtained from SET cotunneling transport measurements. (Triangles) The splittings of the ZBP in the short constriction ($V_B=V_T=0$). (Squares) The splittings of the ZBP in the long constriction ($V_B=V_T= -622$ mV). (Diamond) Kondo splitting at the mid-point of CB valley extracted from figure \ref{fig:splitting_kondo}(b).}
\end{figure}

The ``smooth", CB-free regime persists over a range of voltages $V_T$, making it possible to compare the ZBPs observed at different aspect ratios of the device potential.  As $V_G$ is scanned, both configurations show  bunching of the non-linear traces at $V_{ds} \sim 2$ mV that occurs as the lowest transport band enters the transport window, as  commonly observed in split-gate point contacts \cite{Kristensen:00,cronenwett:02,luscher:07}. The ZBP, clearly seen in the data, splits when the magnetic field is applied (Fig. \ref{fig:data_vs_gate}). We define the peak splitting $\Delta/e$  as half the separation between the two peaks. We note that the separation between the two peaks in bias voltage for a conventional spin 1/2  Kondo effect is expected to be approximately twice the ratio of the Zeeman energy to the electron charge ($\Delta_Z/e$) , and the same is true for the voltage difference between the spin-flip cotunneling steps which we use to measure the g-factor. It is thus interesting to compare $\Delta/e$ to $\Delta_Z/e$ when the CB is not present. First, we note that  $\Delta/e$ shows no strong dependence on the gate voltage. Apart from relatively small departures \cite{Amasha:05,liu:09}, a similar behavior is expected for the conventional Kondo effect. Representative data at  B= 9 T for the short and B= 7 T for the long constrictions are shown in Fig. \ref{fig:data_vs_gate}. The behavior of the ZBP in both regimes is similar: as the device becomes more open, the two peaks become less defined, however, the splitting stays close to the values expected from the Zeeman energy data, marked on the plots with the dashed lines. Next, we fix the gate voltage and focus on the dependence of the splitting on the magnetic field.  Figure \ref{fig:split} shows the comparison of the peak splitting  ($\Delta/e$) to the Zeeman splitting ($\Delta_{Z}/e$).  We find that the splitting of the ZBP increases approximately linearly with the field, and  follows closely the Zeeman splitting data obtained from cotunneling measurements. For comparison, we also show the peak splitting obtained from the data shown in Fig. \ref{fig:splitting_kondo}(b), when Coulomb blockade is present. As expected at large $B$, for the conventional Kondo effect, we find a value which is slightly larger than $\Delta_Z/e$.\cite{liu:09}

The response of the ZBP in QPCs to an in-plane magnetic field is presently not understood. Several groups reported a ZBP that splits at channel conductances comparable to $2e^2/h$ but not at lower conductances \cite{Sarkozy:09,cronenwett:02}.  Chen $et$ $al.$ \cite{chen:09} reported ZBPs in QPCs that did not split with the magnetic field at all, and concluded that the phenomenon did not involve spin. The magnetic splitting of the ZBP  significantly larger than the bulk GaAs Zeeman energy was reported earlier \cite{cronenwett:02}, and attributed  to the enhancement  of the g-factor in one-dimensional conductors \cite{Thomas:98,koop:07,martin:08, chen:09_2}. Such enhancement of the g-factor in open channels has been reported in several experiments: Thomas $et$ $al.$ \cite{Thomas:98} found the effective g-factor enhanced from 0.4 to $\sim 1.2$. Koop $et$ $al.$ \cite{koop:07} found a g-factor enhanced by as much as a factor of $\sim 3$ as compared to  the bulk material, and  a very recent work  \cite{chen:09_2} also reported the enhanced g-factor in an open channel as well as its dependence on carrier density. Compared to these observations, our measurements are performed at relatively low (less than $e^2/h$ ) conductance values, i.e. in the tunneling regime when no actual one-dimensional channel is formed. This may explain the absence of a similar g-factor enhancement in our data. Importantly, the peak splitting we report is in a regime where no signatures of Coulomb blockade ( no conductance ``diamond") with  a conventional Kondo effect are present, and also a direct comparison between the ZBP splitting and the Zeeman energy measured on the same chip is possible.
 
A geometry that favors a formation of a bound state inside the channel was used previously by Sfigakis $et.$ $al.$ \cite{sfigakis:08}, who performed extensive measurements of the temperature dependence of the zero-bias anomaly and concluded that the 0.7 structure and the singlet Kondo effect in a wire are two distinct effects. In that work, a two-gate geometry with small extensions near the ends of the contact was used, and, therefore,  localization of an electron and the  overall transmission of the channel were controlled by the same gate voltage. Thus, in the low conductance regime that we focus on in this work, it was not possible to independently control the degree of the confinement of the electron and the channel conductance. By contrast, the four-gate geometry used in our device allows to controllably create or eliminate the bound state even at the very low conductance values. Our findings suggest that spin-dependent phenomena influence QPC transport even when the tunneling is relatively weak, in addition to the near-opening regime studied extensively by other groups recently \cite{cronenwett:02,luscher:07,sfigakis:08}.

In summary, we have observed a good quantitative agreement between the electron Zeeman energy and the magnetic splitting of a ZBP in a quantum point contact  at conductance values significantly below the first plateau. This result is robust with respect to moderate distortions of the longitudinal potential of the QPC achieved via additional gates in the device design, and shows that even a relatively weak tunneling current in a QPC may be influenced by  spin-dependent effects. Significant distortions of the potential produce a conventional bound charge state accompanied by  the Coulomb blockade and Kondo transport features similar to those found in quantum dots. Coulomb blockade behavior is not present when the QPC potential is smooth. This suggests that an accidental trapping of charge in the channel and the ensuing singlet Kondo effect as observed in quantum dots is not the origin of the ZBP observed in our sample, even though the behavior of our ZBP in the magnetic field and that of a Kondo peak is very similar: both exhibit splitting in bias voltage which is close to twice the Zeeman energy divided by the electron charge and is approximately independent of the gate voltage.

The authors thank M. Jarrell, R. Serota and M. Ma for helpful discussions, and A. Maharjan and M. Torabi for their help with the circuit construction, and J. Markus, M. Ankenbauer and R. Schrott for the technical assistance. T.-M. L. acknowledges device fabrication support from the Institute for Nanoscale Science and Technology at University of Cincinnati. The research is supported by the NSF DMR Award No. 0804199 and by University of Cincinnati.

\mbox{}
\thispagestyle{empty}
\newpage
\mbox{}
\thispagestyle{empty}
\newpage
\mbox{}
\thispagestyle{empty}
\newpage

\mbox{}
\thispagestyle{empty}
\newpage

\mbox{}
\thispagestyle{empty}
\newpage

\begin{figure}
\includegraphics[width=5in, keepaspectratio=true]{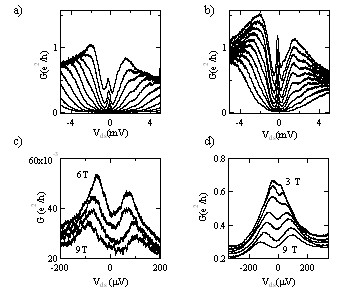}
\caption{\label{fig:map} (a), (b) Observed ZBPs for the short constriction: $V_G$ from -1735 (bottom) to -1697 mV (top) and long constriction: $V_G$ from -1512 (bottom) to -1484 mV (top) at zero magnetic field. Data shown in the figure are obtained by scanning $V_G$ only. (c), (d) Magnetic field dependence of the peak shape for the two configurations, showing that the splitting increases with the  field in both cases. The traces are taken at $V_G=-1630 mV$ (short) and $V_G=-1490 mV$ (long).}
\end{figure}
\mbox{}
\thispagestyle{empty}
\newpage

\end{document}